\documentclass[runningheads]{llncs}

\usepackage{graphicx}
\usepackage{placeins}
\usepackage{amssymb}
\usepackage{amsmath}
\usepackage{url}
\usepackage{xcolor}
\usepackage{verbatim} 
\usepackage{adjustbox} 
\usepackage{multirow} 
\usepackage{hyperref}
\usepackage{booktabs}
\usepackage{colortbl}
\usepackage{subcaption}
\usepackage{stmaryrd}
\usepackage{textcomp}
\usepackage{algorithm}
\usepackage{algpseudocode}
\usepackage{adjustbox}
\usepackage{rotating}

\newcommand{\task}{\mathcal{T}}

\newcommand{\memory}{\mathcal{M}}
\newcommand{\expectation}{\mathbb{E}}

\definecolor{tabbestcolor}{rgb}{0.004, 0.141, 0.337}

\def \best {\cellcolor{tabbestcolor!30}}
\def \sbest {\cellcolor{tabbestcolor!15}}

\algnewcommand\algorithmicforeach{\textbf{for each}}
\algdef{S}[FOR]{ForEach}[1]{\algorithmicforeach\ #1\ \algorithmicdo}

\begin{document}
\title{Distribution-Aware Replay for Continual MRI Segmentation}

\author{Nick Lemke\inst{1} \and
Camila Gonz\'alez\inst{2} \and
Anirban Mukhopadhyay\inst{1} \and \\
Martin Mundt\inst{1,3}}
\authorrunning{N. Lemke et al.}
\institute{Technical University of Darmstadt, Darmstadt, Germany
\and
Stanford University, Stanford, USA
\and
The Hessian Center for Artificial Intelligence: hessian.AI, Darmstadt, Germany
\email{nick.lemke@gris.informatik.tu-darmstadt.de}}

\maketitle              
\begin{abstract}
    Medical image distributions shift constantly due to changes in patient population and discrepancies in image acquisition. These distribution changes result in performance deterioration; deterioration that continual learning aims to alleviate. However, only adaptation with data rehearsal strategies yields practically desirable performance for medical image segmentation. Such rehearsal violates patient privacy and, as most continual learning approaches, overlooks unexpected changes from out-of-distribution instances. To transcend both of these challenges, we introduce a distribution-aware replay strategy that mitigates forgetting through auto-encoding of features, while simultaneously leveraging the learned distribution of features to detect model failure. We provide empirical corroboration on hippocampus and prostate MRI segmentation. To ensure reproducibility, we make our code available at \url{https://github.com/MECLabTUDA/Lifelong-nnUNet/tree/cl_vae}.
\end{abstract}

\section{Introduction}
Deep learning approaches are largely regarded as successful in static biomedical image segmentation settings~\cite{nnunet}. Yet, medical data may shift according to changes in the patient population, vary according to disease-related factors, or be subject to differences resulting from nuances in image acquisition parameters~\cite{sahiner2023data}. Since medical image segmentation models are typically trained on small datasets (judged by deep learning standards), they tend to not generalize well to such \emph{shifted distributions}~\cite{geirhos2020shortcut}. Ideally, a learner should be able to expand its knowledge by training on new samples from the prospectively shifted or later recorded distributions. As do medical experts, our artificial system should \emph{learn continually}~\cite{Mundt2023wholistic}. In order to enable the latter it is required to overcome a phenomenon understood as \emph{catastrophic forgetting}~\cite{McCloskey1989,Ratcliff1990}, or more intuitively, to avoid new information from greedily overwriting existing knowledge.    

However, in medical imaging, continual learning algorithms are so far not the remedy that was promised. Among the conceptual pillars of proposed algorithms \cite{Parisi2019}, rehearsal of data subsets \cite{Rolnick2018} performs by far the best, yet directly violates inherent (medical) \emph{privacy} regulations~\cite{GDPR_EU}.
Generative replay~\cite{shin2017continual} aims at capturing the distributions encountered during training, and including synthesized data in future training tasks. However, compared to distributions of natural images, those of MRIs are much more difficult to grasp as MRIs are more complex and more high-dimensional.
Alternative methods that instead rely on constraining model parameters, so-called regularization approaches \cite{kirkpatrick2017overcoming,zenke2017continual}, have in turn been shown to perform poorly on medical data~\cite{gonzalez2020wrong}. In fact, this failure mode of forgetting due to having no access to past data is further exacerbated by an often overlooked additional phenomenon - the \emph{silent failure} of models. They not only suffer from expected forgetting of past experiences, but also produce overly confident false predictions whenever unexpected data is encountered~\cite{Boult2019}. Again ideally, the learner should be able to detect and outright reject these \emph{out-of-distribution} (OoD) examples. Unfortunately, the latter is substantially challenged by the reality that predominant segmentation models like UNet~\cite{UNet,nnunet} lack a notion of the learned distribution. Existing OoD detection algorithms thus often assume a-priori knowledge of the anticipated OoD samples~\cite{dhamija2018reducing,Lee2018} or hope that expensive uncertainty approximations capture the examples~\cite{kendall2015bayesian,liang2017enhancing}. On the contrary, generative models~\cite{kingma2013auto} (that explicitly learn the distribution) are notoriously hard to train for discriminatory tasks.   

In this work, we simultaneously address the challenge of avoiding forgetting without direct violations of privacy in continual learning and overcome silent prediction failures by rejecting OoD instances. To this end, we leverage prior insights on theoretically grounded two-stage modeling~\cite{dai2018diagnosing,hong2022return}, where a second generative model encodes the distribution of our primary discriminative model, without interfering in the latter's learning or inference processes. Specifically, we propose a second-stage conditional variational autoencoder (VAE)~\cite{kingma2013auto} to model the low-dimensional distribution of a UNet's latent features. With the feature distribution captured by the VAE we can then make rigorous decisions to assess whether a new subject is outside the known distribution and conversely employ a pseudo-rehearsal setup to replay features of past subjects to avoid forgetting when adapting the model continually. We evaluate our setup on domain incremental MRI segmentation tasks of the hippocampus and the prostate and further assess the OoD detection capabilities on augmented datasets. 
\section{Methodology}
\begin{figure}[t]
    \centering
    \includegraphics[width=\textwidth]{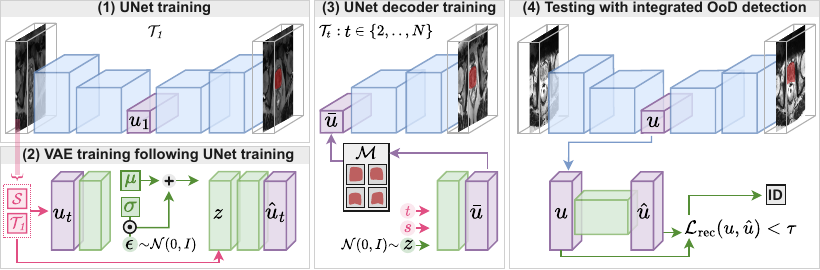}
    \caption{ 
    (1) The UNet is trained on the first task $\task_1$. (2) The VAE is trained on features $u_1$ with slice and task conditioning. 
    (3) A set of features $\bar u_{i<t}$ are synthesized, pseudo-labeled and placed in memory $\memory$. The UNet decoder is then trained on $\memory$ and the new data of task $\task_t$. 
    (4) During inference, the reconstruction loss between $u$ and $\hat u$ is used to classify whether the MRI is ID or OoD.}
    \label{fig:unet_vae}
\end{figure}

The UNet architecture~\cite{UNet} is well-known for its extraordinary performance in medical image segmentation~\cite{nnunet}. How do we leverage this architecture and retain its efficacy while overcoming its inherent forgetful nature and its silent failure modes? To achieve symbiosis between these desiderata we leverage recent theoretical insights~\cite{dai2018diagnosing}, proving that a second VAE can correctly model an initial VAE's learned distribution as an isotropic Gaussian distribution as a consequence of the known hidden dimensionality of the first model. This in turn allows to replay the learned distribution in continual learning~\cite{hong2022return}. As we will proceed to elaborate, placing such a VAE meta-model on top of a medical UNet will now allow us to i) model and rehearse the feature distribution of a UNet without interfering in its learning process, ii) strategically condition the VAE on observed tasks and volumetric slicing of the medical data, iii) leverage the represented feature distribution to reject OoD examples to avoid silent model failure. Fig.~\ref{fig:unet_vae} shows a schematic representation of the proposed architecture. 

\subsection{A Two-stage Architecture for Continual Medical Segmentation}
Consider a UNet composed of several blocks of convolutions to downsample the data and then recombine the representation to produce a segmentation map. Conceptually, a UNet is comprised of an encoder, encoding the features of the data into a latent code $u$, followed by a decoder, decoding the code into the desired output. However, as the model is trained in a supervised discriminative fashion, we unfortunately do not know the form of the distribution of $u$. We overcome this hurdle by capturing $p(u)$ through a separate VAE. The goal of this model is to learn an approximate posterior $q(z|u)$ through variational inference, where $z$ is a second set of latent factors which we optimize to follow a pre-defined prior $p(z)$. This prior is an easy-to-sample Normal distribution. The key is that the latent code $z$ has the same dimensionality as $u$. Thus, we can encourage the VAE to learn a lossless mapping from our UNet's unknown feature distribution $p(u)$ to our prior with the aid of a decoder that models the likelihood $p(u|z)$. We can then train the VAE with an evidence lower bound:
$\log p(u) \geq \expectation_{z\sim q(z|u)}\left[\log p(u|z)\right] - \text{KL}\left[q(z|u)||p(z)\right]$. Here KL denotes the Kullback-Leibler divergence. The UNet training is shown in Fig.~\ref{fig:unet_vae} (1), followed by the VAE training after each UNet update step in Fig.~\ref{fig:unet_vae} (2). On arrival of a new task $\task_{t>2}$ a buffer $\memory$ of pseudo-samples is synthesized by the VAE posterior and pseudo-labeled by the latest UNet decoder. The pseudo-elements and the data from the new task are used to update the UNet decoder as shown in Fig.~\ref{fig:unet_vae} (3). At the end of the training loop, the VAE is updated using the same memory buffer and the new data (Fig.~\ref{fig:unet_vae} (2)).

\subsection{Distribution-aware Pseudo-replay with Native OoD Detection}
Intuitively, our UNet first trains on a task $\task_1$ (Fig. \ref{fig:unet_vae} (1)) and subsequently the VAE learns to model the encoded feature distribution (Fig. \ref{fig:unet_vae} (2)). In principle this already allows us to 1) assess whether new samples are dissimilar to already observed ones, 2) rehearse previous experience by generating pseudo-data~\cite{Robins1995}. However, to adequately maintain knowledge of each task we have observed in continual learning, we further condition our VAE on the task identity $t$, i.e. $t$ is appended to the VAE input $u$ and the latent variable $z$. As the learned task embedding encodes the unique properties of each domain, the VAE remains fixed in size as more distributions are captured.

This conditioned VAE entails multiple advantages. For the above first ability, OoD detection, it enables us to use the VAE's predicted log-likelihood (the reconstruction loss) to decide whether a new sample during UNet inference is dissimilar to any previous tasks' distribution. Once the VAE observes more than one task, we consider the lowest reconstruction error obtained with each previous task identity $t$. Specifically, we classify samples with a reconstruction error below a threshold $\tau$ as in-distribution (ID), which we calibrate on the $95\%$ true positive rate on the validation set (Fig. \ref{fig:unet_vae} (4)). Importantly, such an OoD detection procedure does not interfere with the UNet's semantic segmentation prediction, maintaining it's well-known precision and merely augmenting it with an OoD score to inform the user of (un-)trustworthy predictions. 

For the second ability, mitigation of forgetting, we use the conditioned VAE to generate pseudo-features $\bar u_{i<t}$ for past experiences in the former sequence of tasks $\task_1,\task_2,\dots$. Here, the task conditioning ensures that we can synthesize a balanced memory $\memory$. Specifically, as we progress through tasks the MRIs are first encoded to features $u$ using the UNet encoder, on which the VAE trains with the additional conditioning. To avoid forgetting of these tasks when proceeding to a new task $\task_{t+1}$, we then fill a memory of synthesized examples by: 1) sampling $z$ for each respective task $z \sim p(z|t)$ from the Normal distribution in our VAE, 2) using its decoder to map this random value to a UNet's pseudo-feature $\bar u$ that is alike previous experience, and 3) inferring the pseudo-feature's label with the UNet decoder (Fig. \ref{fig:unet_vae} (3)).  To ensure that the distribution of features does not change as we continue training the decoder, we freeze the UNet's encoder after the first task. Finally, after each task's training, the encoded features of the UNet are then deleted, and the current memory is flushed to reduce the memory footprint and ensure adherence to privacy considerations.

\subsection{MRI Advantages Through VAE Double Conditioning}
Following theory~\cite{dai2018diagnosing}, the distribution is only correctly learned by the VAE if its latent dimension matches the UNet encoder's feature dimensionality: $\mathtt{dim}(z) == \mathtt{dim}(u)$. Though already low-dimensional, our 3D UNet still has a spatial resolution of $5\times7\times5$ with $256$ channels. This results in a latent space of size $5\times7\times5\times256=44,800$, which remains cumbersome. To make our final model computationally feasible, we restrict the UNet to be two-dimensional by segmenting slice-wise along the lowest resolution, reducing the dimension by a factor of $5$ to $8,960$. The two-dimensional UNet is thus applied to slices of the 3D image volume and the smaller latent space is well learnable by the VAE. However, we now expect large differences in the features between slices at different locations in the volume. To ensure that this choice does not become detrimental, we introduce a final conditioning into the VAE: a further slice index $s$ to indicate the position of the slice within the volume. We refer to this doubly-conditioning architecture as \textbf{ccVAE} and show its empirical superiority in the following.

\section{Experimental Setup}
\begin{figure}[]
    \centering
    \begin{subfigure}{\textwidth}
    \centering
    \includegraphics[width=\textwidth]{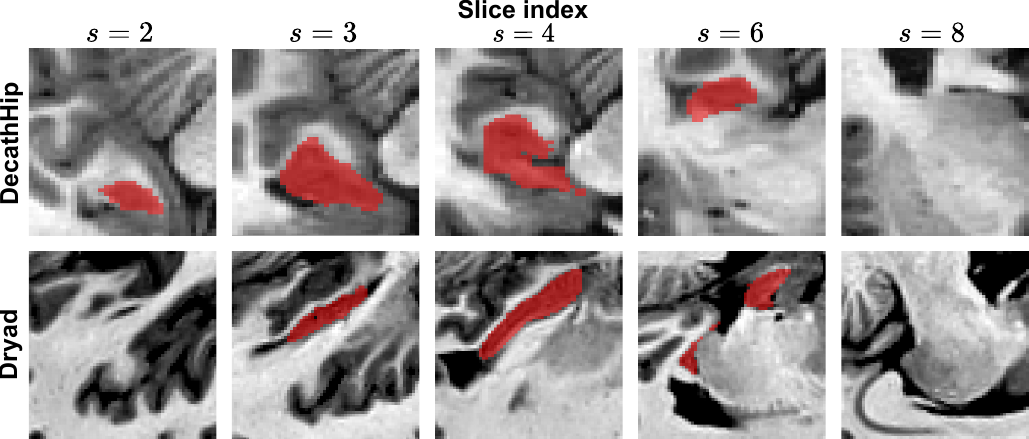}
    \caption{}
    \end{subfigure}
    \begin{subfigure}{\textwidth}
    \centering
    \includegraphics[width=\textwidth]{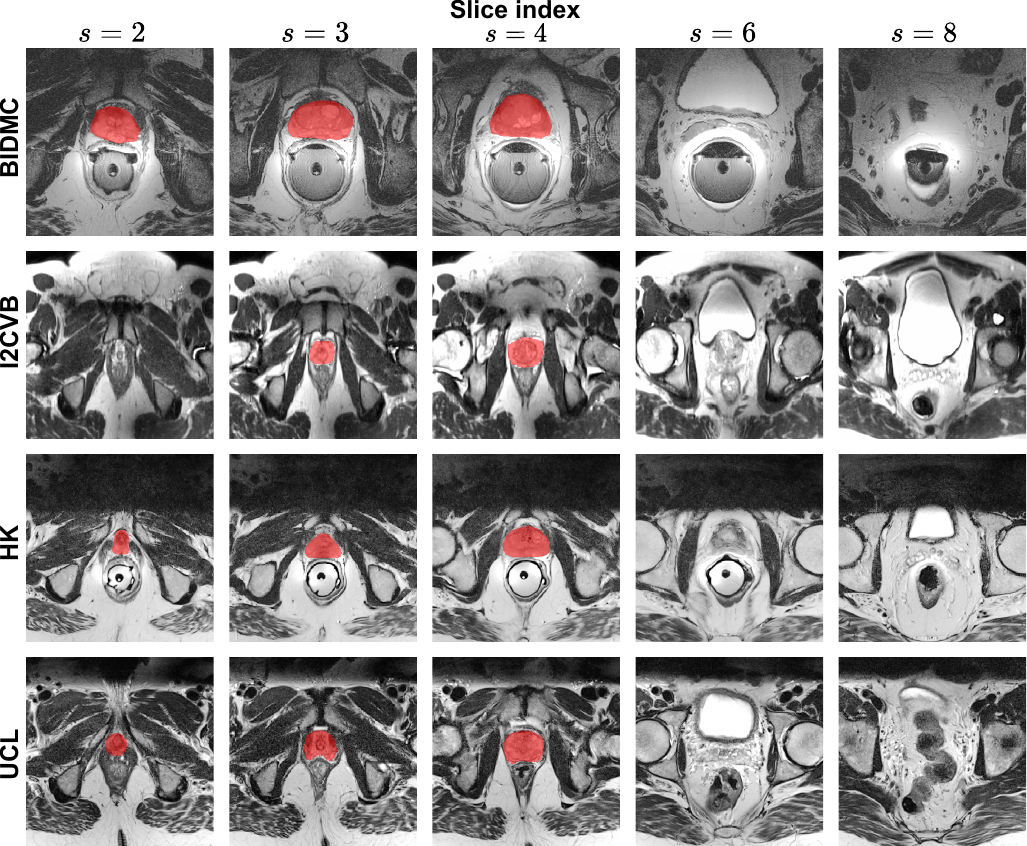}
    \caption{}
    \end{subfigure}
    \caption{Representative slices $s$ of MRI scans from each (a) hippocampus and (b) prostate dataset. The red areas depict the ground truth segmentation masks.}\label{fig:qualitative_data_comparison}
\end{figure}

\begin{figure}
    \centering
    \includegraphics[width=\textwidth]{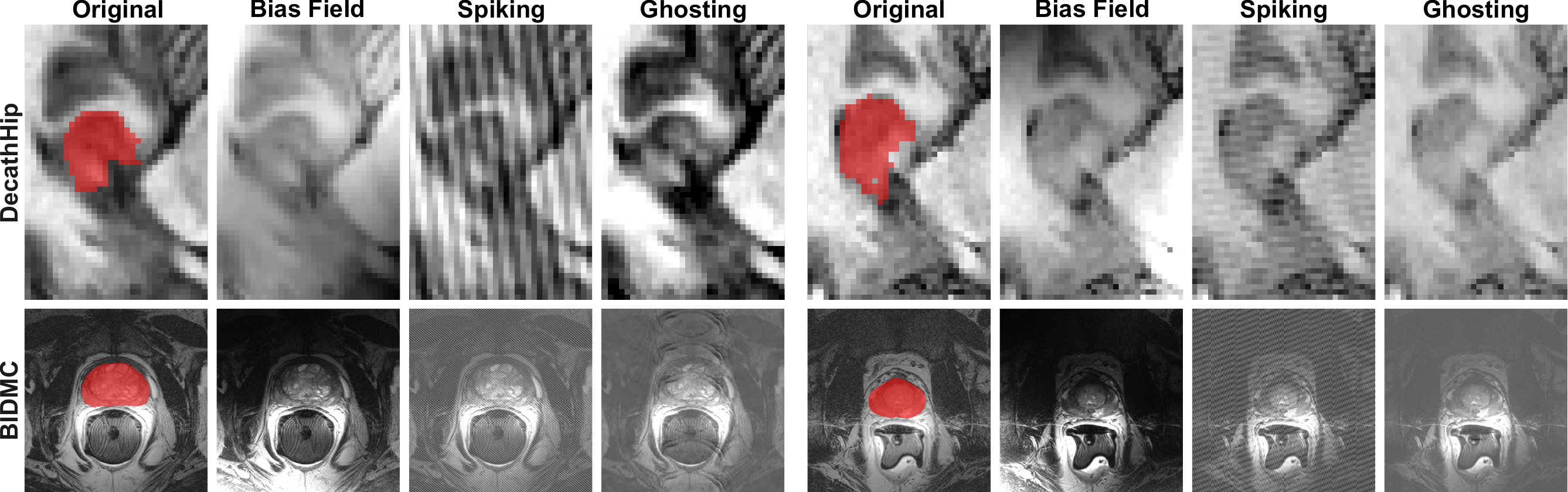}
    \caption{Augmentations applied to the hippocampus (top row) and prostate (bottom row) datasets to create challenging OoD scenarios.}
    \label{fig:data_augmentations}
\end{figure}

\noindent \textbf{Data:}
Following previous work on medical continual segmentation \cite{gonzalez2022task,gonzalez2023lifelong,ranem2024continual,ranem2022continual}, we evaluate on the tasks of segmenting the prostate and hippocampus in, respectively, T2-weighted and T1-weighted MRIs. The \textbf{hippocampus} data consists of three datasets: \emph{Multi-contrast submillimetric 3 Tesla hippocampal subfield segmentation} (\emph{Dryad}) \cite{kulaga2015multi}, \emph{Harmonized Hippocampal Protocol dataset} (\emph{HarP})~\cite{wisse2017harmonized} and the
hippocampus data released for the \emph{Medical Segmentation Decathlon} (\emph{DecathHip})~\cite{antonelli2022medical}. We train in the order \emph{DecathHip}$\rightarrow$\emph{Dryad} following the setup in previous works~\cite{gonzalez2020wrong}. We preserve \emph{HarP} for OoD testing. The sets contain 260, 50, and 270 samples, respectively. The \textbf{prostate} data originates from five institutions using different devices and acquisition parameters~\cite{liu2020ms}. We train in the order \emph{BIDMC}$\rightarrow$\emph{I2VCB}$\rightarrow$\emph{HK}$\rightarrow$\emph{UCL}, creating a challenging setting by starting with the smallest dataset and alternating between datasets with and without an endorectal coil. The segmentation mask encompasses the central gland and peripheral area. We likewise use the final dataset, \emph{RUNMC} for OoD evaluation. Each dataset contains 12 to 30 samples and is randomly divided into $20\%$ testing, $56\%$ training, and $24\%$ validation. A qualitative comparison of the data used can be found in Fig.~\ref{fig:qualitative_data_comparison}. We also utilize synthetic OoD data. Here, we augment the test sets with common MRI artifacts (random bias field, spiking, or ghosting) doubling their size. A few examples of augmented MRIs are depicted in Fig.~\ref{fig:data_augmentations}.\\

\noindent \textbf{Architectures and Training:}
We use the state-of-the-art nnUNet framework~\cite{nnunet}, which automatically configures UNet parameters based on data characteristics. Our VAE consists of 8 linear layers with batch norm and leaky ReLU, and is trained for 5000 epochs. We use the Adam optimizer~\cite{kingma2014adam}, an initial learning rate of $1e-3$, and exponential learning rate decay with a rate of $9.9e-1$. For generated features there are no activations in the UNet encoder, so we discard the skip connections. We run our experiments on two Nvidia A40 GPUs.\\

\addvspace{4mm}
\noindent \textbf{Baselines:}
We compare ccVAE to regular sequential (\emph{Seq.}) training and several continual learning methods with comparable privacy preservation. Elastic weight consolidation (\emph{EWC})~\cite{kirkpatrick2017overcoming} penalizes the deviation of parameters deemed to be significant for past tasks. Modeling the background (\emph{MiB})~\cite{cermelli2020modeling} is tailored specifically for semantic segmentation and uses an unbiased distillation loss that penalizes a shift in the foreground classes.
For OoD detection during testing, we use the maximum softmax probability (\emph{SM})~\cite{hendrycks2016baseline}.
We also compare to maintaining a pool of models trained at different stages (\emph{MPool})~\cite{gonzalez2022task} and using Segmentation Distortion (\emph{SD})~\cite{lennartz2023segmentation} for OoD detection, which similarly to our approach uses an autoencoder for reconstructing features of a pre-trained UNet.
During inference, the UNet corresponding to the autoencoder with the lowest SD is chosen for segmentation. Finally, we do an ablation of ccVAE by using only conditioning on the task (\emph{cVAE}) and detecting OoD samples based on the Mahalanobis distance in the feature space (\emph{Mah}) \cite{gonzalez2022distance} instead of the reconstruction error.\\

\noindent \textbf{Metrics:}
For evaluating the segmentation performance of continually trained models, we compute the Dice score for the samples classified as in-distribution. We also report the expected calibration error (ECE)~\cite{guo2017calibration} after normalization, as well as the backward (BWT) and forward (FWT) transferability ~\cite{gonzalez2023lifelong}.
\section{Results}
We first evaluate ccVAE in a challenging setting with abrupt shifts in the data distribution during continual training. We further introduce OoD data during testing, first in the form of an unseen dataset and later by adding image artifacts.\\

\begin{figure}[]
    \centering
    \begin{subfigure}{.5\linewidth}
        \centering
        \includegraphics[width=\linewidth]{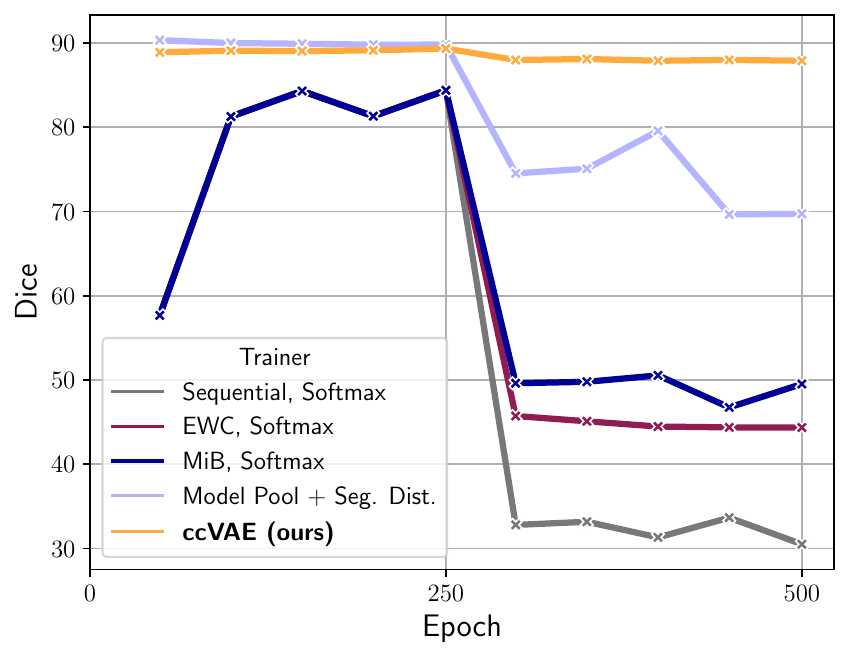}
        \caption{Hippocampus}
        \label{fig:ood_and_cl:hippocampus}
    \end{subfigure}%
    \begin{subfigure}{.5\linewidth}
        \centering
        \includegraphics[width=\linewidth]{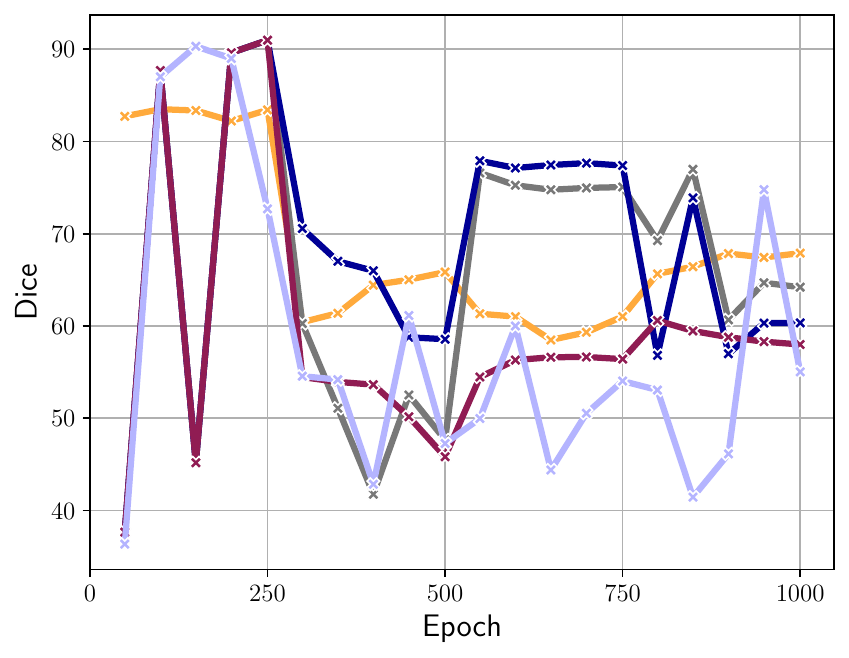}
        \caption{Prostate}
        \label{fig:ood_and_cl:prostate}
    \end{subfigure}
    \caption{Test Dice ($\uparrow$) during the learning trajectory for (a) hippocampus and (b) prostate. New tasks are introduced at 250 epoch intervals. ccVAE (yellow) maintains the most stable segmentation performance throughout the trajectory.}
    \label{fig:ood_and_cl}
\end{figure}

\noindent \textbf{Continual Learning Under Dataset Shift:} Fig.~\ref{fig:ood_and_cl} illustrates the performance of ccVAE alongside existing methods in a continual learning context, where new tasks are introduced at 250 epoch intervals. The y-axis depicts the mean Dice for test images from all tasks that are considered ID. After the shift in the hippocampus data, only ccVAE learns to adapt while still producing high-quality segmentations for the older distribution, consequently maintaining robust performance across the trajectory. The expansion-based pooling baseline with segmentation distortion also remains mostly unaffected by the shift but is outperformed by ccVAE. Continual segmentation of the prostate proves more challenging. There is an abrupt fall in segmentation quality after the second task is introduced, likely due to the small size of the database (7 to 11 samples per task) that makes generalization more challenging. As ccVAE recognizes samples from more than the present task as ID and attempts to segment them, we see the performance on $\task_1$ deteriorate. However, from that point on, ccVAE remains stable while other methods display noticeable volatility in segmentation performance.

\begin{table}[]
    \centering
    \setlength\tabcolsep{2pt}
    \begin{tabular}{l|rrr|rrr}
\toprule
Anatomy/&\multicolumn{3}{c|}{Hippocampus}&\multicolumn{3}{c}{Prostate}\\
                              Method &  Dice $\uparrow$&   BWT $\uparrow$&  FWT $\uparrow$&  Dice $\uparrow$&   BWT $\uparrow$&  FWT $\uparrow$\\
\midrule
           Sequential 									 & 20.1$\pm$32.1 &  -83.2$\pm$8.2 &    \best0.0$\pm$0.0 		& 54.7$\pm$30.9 & -43.3$\pm$29.6 &    \sbest0.0$\pm$0.0 			\\
                  EWC \cite{kirkpatrick2017overcoming} 	 & \sbest77.5$\pm$28.0 &   \best0.0$\pm$0.2 &  -77.3$\pm$6.3 	& 53.5$\pm$28.8 &    \best2.1$\pm$8.6 & -47.0$\pm$28.4 			\\
                  MiB \cite{cermelli2020modeling} 		 & 60.6$\pm$16.9 & -34.9$\pm$10.7 &   \sbest-1.1$\pm$0.8 		& 53.3$\pm$32.0 & -45.6$\pm$27.9 &    \best0.4$\pm$3.4 			\\
                MPool \cite{gonzalez2022task} 			 & 72.8$\pm$33.0 & -13.2$\pm$31.5 & -37.5$\pm$36.3 			& \sbest54.8$\pm$35.0 &   \sbest0.9$\pm$41.6 & -44.1$\pm$35.6 	\\
\textbf{ccVAE (ours)} 									 &  \best87.8$\pm$4.5 &   \sbest-1.3$\pm$4.8 &   -3.9$\pm$2.0 	&  \best64.5$\pm$9.1 & -11.4$\pm$10.1 &  -17.0$\pm$7.7 			\\
\bottomrule
\end{tabular}
    \caption{Mean Dice, backward transfer (BWT) and forward transfer (FWT) of the model for all test samples after training on the hippocampus and prostate sequences, respectively. ccVAE achieves the best segmentation performance, with little forgetting and robust knowledge accumulation.}
    \label{tab:bwt_fwt}
\end{table}

\noindent Tab.~\ref{tab:bwt_fwt} reports the average Dice, BWT and FWT after the entire training sequence, regardless of whether samples are considered ID or OoD. Sequential training and MiB suffer from substantial forgetting, shown by a large negative BWT and overall lower Dice scores. The expansion-based MPool successfully prevents forgetting, yet at the cost of a loss in plasticity as most members from the model pool do not acquire knowledge from the latter training stages.\\

\noindent\textbf{Navigating Dataset Shift and Image Artifacts:} We now increase the difficulty of the training conditions further by augmenting the test images with synthetically generated MRI artifacts. Table~\ref{tab:augmentation_res} shows the Dice of all images deemed to be ID, alongside the expected calibration error calculated on all test samples. We report the results after each training stage. ccVAE consistently performs well in early stages, showing its ability to identify cases that it can segment successfully. All methods struggle after training with \emph{HK} (column 5), which proves particularly challenging. Here, sequential and MiB training perform well in a trade-off that only considers images from the latest task as ID, disregarding the earlier tasks. As they are both highly plastic methods, they quickly adapt to this new task. ccVAE, on the other hand, considers most images following distributions seen in the past as ID. This demonstrates that despite having some protection against forgetting in the form of generated pseudo-samples, a highly shifted dataset in the sequence will damage the segmentation ability. Still, performance of ccVAE across the trajectory and within each evaluation round remains stable, as corroborated by the consistently low standard deviation in ccVAE predictions. \\

\begin{table}[]
    \centering
    \adjustbox{max width=\textwidth}{%
    \begin{tabular}{l|cc|cc||cc|cc|cc|cc}
\toprule
Training stage/&\multicolumn{2}{c|}{\emph{DecathHip}}&\multicolumn{2}{c||}{\emph{Dryad}}&\multicolumn{2}{c|}{\emph{BIDMC}}&\multicolumn{2}{c|}{\emph{I2CVB}}&\multicolumn{2}{c|}{\emph{HK}}&\multicolumn{2}{c}{\emph{UCL}}\\
Method &Dice $\uparrow$&\textbf{E} $\downarrow$&Dice $\uparrow$&\textbf{E} $\downarrow$&Dice $\uparrow$&\textbf{E} $\downarrow$&Dice $\uparrow$&\textbf{E} $\downarrow$&Dice $\uparrow$&\textbf{E} $\downarrow$&Dice $\uparrow$&\textbf{E} $\downarrow$\\
\midrule
                                 Seq., SM \cite{hendrycks2016baseline}  &       63.4$\pm$39 &       51.1 &       19.4$\pm$31 &       48.3 & \sbest50.5$\pm$40 &       39.8 &       38.8$\pm$36 &       40.3 &  \best71.0$\pm$16 & \sbest26.7 &  \best58.9$\pm$28 &  \best16.7\\
 EWC \cite{kirkpatrick2017overcoming}, SM \cite{hendrycks2016baseline}  &       63.4$\pm$39 &       51.1 &       32.6$\pm$38 &       49.6 & \sbest50.5$\pm$40 &       39.8 &       37.3$\pm$32 &       34.2 &       46.2$\pm$27 &       30.2 &       48.2$\pm$26 & \sbest25.3\\
      MiB \cite{cermelli2020modeling}, SM \cite{hendrycks2016baseline}  &       63.4$\pm$39 &       51.1 &       26.5$\pm$31 &       45.3 & \sbest50.5$\pm$40 &       39.8 & \sbest44.3$\pm$30 &  \best20.6 & \sbest70.7$\pm$16 &  \best21.8 &       48.5$\pm$33 &       31.8\\
      MPool \cite{gonzalez2022task}, SD \cite{lennartz2023segmentation}  & \sbest82.4$\pm$24 & \sbest48.3 & \sbest47.8$\pm$40 & \sbest42.4 &       47.2$\pm$42 & \sbest37.2 &       37.6$\pm$34 &       43.4 &       46.4$\pm$34 &       37.2 &       41.4$\pm$36 &       34.4\\
                                                 \textbf{ccVAE (ours)}  &  \best89.3$\pm$ 3 &   \best7.8 &  \best83.2$\pm$14 &   \best4.7 &  \best75.6$\pm$11 &  \best14.8 &  \best56.7$\pm$17 & \sbest21.5 &       49.4$\pm$21 &       27.8 & \sbest58.8$\pm$15 &       32.3\\
\bottomrule
\end{tabular}}
    \vspace*{0.5em}
    \caption{Dice for subjects classified as ID and expected calibration error (\textbf{E}CE) after each training stage for all the test data, including cases from each task as well as scans augmented with MRI artifacts. Except for HK, where Seq. SM and MPool trade-off performance, ccVAE demonstrates superior stable performance.}
    \label{tab:augmentation_res}
\end{table}

\begin{figure}[h]
    \centering
    \begin{subfigure}{.25\linewidth}
        \centering
        \includegraphics[width=.95\linewidth]{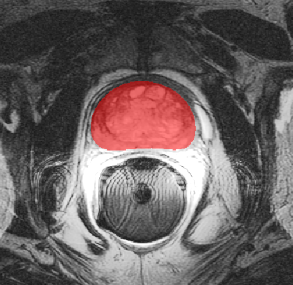}
        \caption{Dice: $86.6\%$}
    \end{subfigure}\hfill
    \begin{subfigure}{.25\linewidth}
        \centering
        \includegraphics[width=.95\linewidth]{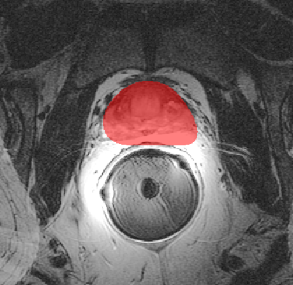}
        \caption{Dice: $81.0\%$}
    \end{subfigure}\hfill
    \begin{subfigure}{.25\linewidth}
        \centering
        \includegraphics[width=.95\linewidth]{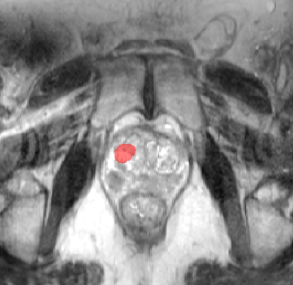}
        \caption{Dice: $11.7\%$}
    \end{subfigure}\hfill
    \begin{subfigure}{.25\linewidth}
        \centering
        \includegraphics[width=.95\linewidth]{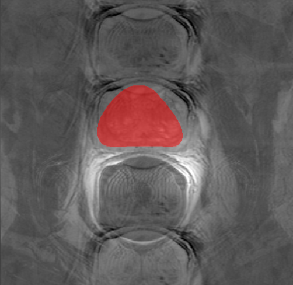}
        \caption{Dice: $73.6\%$}
    \end{subfigure}
    \caption{Four segmentations produced by the model trained on the first prostate dataset. Images (a) and (b) are correctly considered ID and segmented correctly. (c) is correctly considered OoD, but (d) is misclassified.}
    \label{fig:qualitative_results}
\end{figure}

\noindent \textbf{Qualitative Evaluation:} Fig.~\ref{fig:qualitative_results} illustrates four exemplary prostate segmentations produced by ccVAE. The first and second images are ID MRIs that are correctly classified as such and segmented well. The third is an OoD MRI that is segmented poorly but rejected by the OoD detection mechanism. The fourth MRI is augmented with a ghosting artifact and not detected.\\

\noindent \textbf{Ablation Study:} 
In Tab.~\ref{tab:ablations} we ablate ccVAE in the simpler scenario without artifact augmentations to corroborate that all elements of our approach are needed. We compare the proposed ccVAE, which detects OoD samples based on the reconstruction error, to estimating the uncertainty from the Mahalanobis distance to the prior distribution $p(z)$ (\emph{ccVAE, Mah.}). We also evaluate a version of the VAE that is only conditioned on the task (\emph{ccVAE, Rec.}). Alongside these ablations, we include the per-stage results of the model pool with segmentation distortion baseline (\emph{MPool SD}), which is closest in performance to ccVAE in Fig.~\ref{fig:ood_and_cl}. In most stages, the full ccVAE is necessary to obtain the highest Dice and the first or second-lowest ECE. The OoD detection strategy based on the Mahalanobis distance fails to calibrate the model in early training, resulting in high ECEs and low Dice scores.
\begin{table}[h]
    \centering
    \adjustbox{max width=\textwidth}{%
    \begin{tabular}{l|cc|cc||cc|cc|cc|cc}
\toprule
Training stage/&\multicolumn{2}{c|}{\emph{DecathHip}}&\multicolumn{2}{c||}{\emph{Dryad}}&\multicolumn{2}{c|}{\emph{BIDMC}}&\multicolumn{2}{c|}{\emph{I2CVB}}&\multicolumn{2}{c|}{\emph{HK}}&\multicolumn{2}{c}{\emph{UCL}}\\
Method &Dice $\uparrow$&\textbf{E} $\downarrow$&Dice $\uparrow$&\textbf{E} $\downarrow$&Dice $\uparrow$&\textbf{E} $\downarrow$&Dice $\uparrow$&\textbf{E} $\downarrow$&Dice $\uparrow$&\textbf{E} $\downarrow$&Dice $\uparrow$&\textbf{E} $\downarrow$\\
\midrule
MPool \cite{gonzalez2022task}, SD \cite{lennartz2023segmentation}  &  \best89.8$\pm$ 3 &      33.4 &       69.7$\pm$35 &       20.1 &      72.3$\pm$34 &       30.3 &       48.6$\pm$34 &       35.1 &       55.1$\pm$31 & \sbest31.8 &      55.9$\pm$34 &       30.2\\
                         ccVAE, Mah. \cite{gonzalez2022distance}  &       89.0$\pm$ 3 &      13.2 &       61.2$\pm$33 &       24.4 &      39.1$\pm$30 &       29.0 &       60.5$\pm$13 &       34.7 & \sbest60.4$\pm$18 &       34.2 & \best67.9$\pm$10 &  \best22.6\\
                                                      cVAE, Rec.  &       89.3$\pm$ 3 &  \best3.8 & \sbest87.6$\pm$ 4 & \sbest16.8 & \best83.4$\pm$ 2 &  \best24.4 & \sbest64.7$\pm$ 9 &  \best19.4 &  \best65.4$\pm$12 &  \best17.3 &      65.4$\pm$10 & \sbest28.6\\
                                           \textbf{ccVAE}  & \sbest89.4$\pm$ 3 & \sbest4.7 &  \best87.9$\pm$ 5 &  \best14.5 & \best83.4$\pm$ 2 & \sbest25.5 &  \best66.2$\pm$ 9 & \sbest27.2 &       60.0$\pm$19 &       35.5 & \best67.9$\pm$10 &       37.8\\
\bottomrule
\end{tabular}}
    \vspace*{0.5em}
    \caption{Ablation study comparing ccVAE to different versions of our method and the best baseline from Fig.~\ref{fig:ood_and_cl}
    Both conditioning and basing OoD detection on VAE reconstructions consistently contribute to performance.}
    \label{tab:ablations}
\end{table}

\begin{figure}
    \centering
    \includegraphics[width=\textwidth]{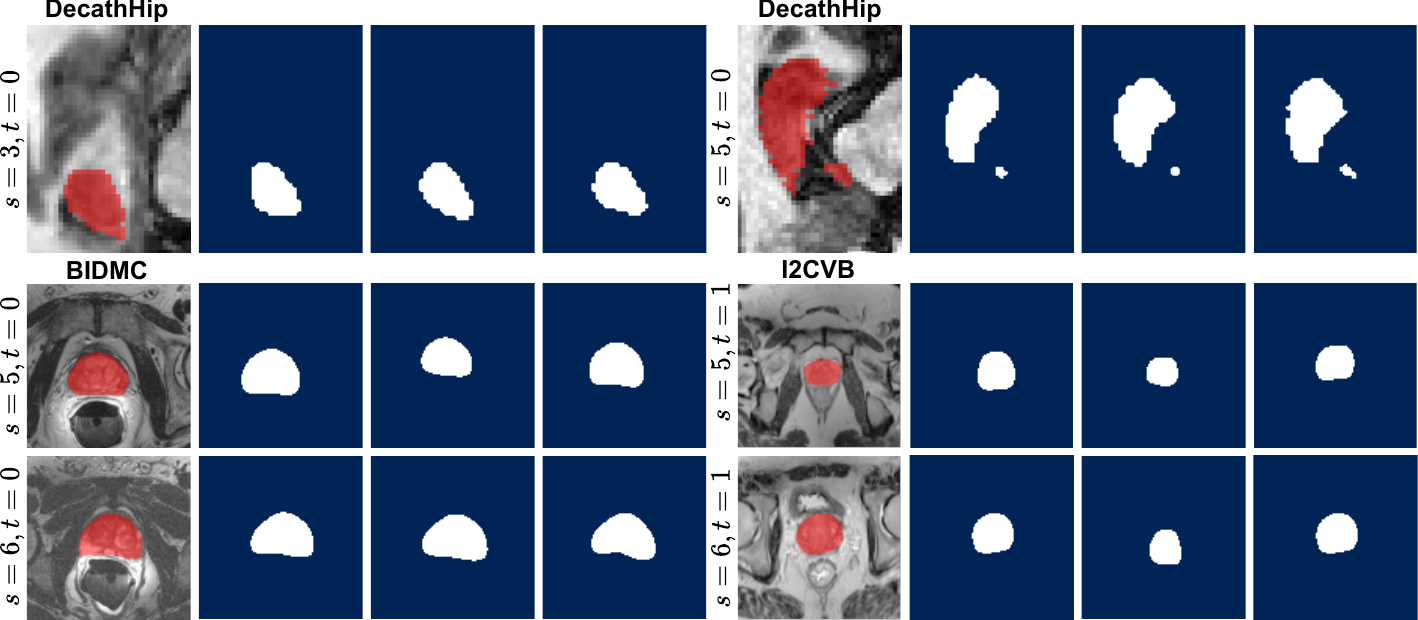}
    \caption{Ground truth segmentation masks from the original tasks and segmentation masks from generated features using different slice and task indices.}
    \label{fig:generated_segmentations}
\end{figure}

\noindent\textbf{Analysis of Generated Features:} Finally, in Figs.~\ref{fig:generated_segmentations} we qualitatively support our quantitative findings by visualizing segmentation masks of the train set and similar segmentation masks of the ccVAE's generated features included in pseudo-rehearsal training. The generated features are semantically coherent, cover multiple volume segments and successfully capture geometric diversity.

\section{Conclusion}

Aiming to translate the success of medical image segmentation to more realistic dynamic settings, where there are abrupt shifts in the training distribution and the model encounters low-quality images during testing, we propose ccVAE. Our method augments UNet segmentation models with a small VAE that maps features into a standard normal distribution without reducing dimensionality. In turn, this allows to generate features similar to those seen in previous tasks, preventing forgetting without compromising patient privacy, and enabling principled OoD detection. ccVAE, therefore, jointly addresses the two main factors causing unexpected performance deterioration in dynamic clinical environments.

\bibliographystyle{splncs04}
\bibliography{main.bbl}
\end{document}